\documentclass[aps,amsmath,amssymb,reprint,prl,superscriptaddress]{revtex4-1}  
\usepackage{graphicx}  
\usepackage{dcolumn}   
\usepackage{bm}        
\usepackage{amssymb}   

\newcommand\diff{\,\mathrm{d}}

\def\sigp{\sigma_{\chi,p}}
\def\sige{\sigma_{\chi,e}}
\def\sigeff{\sigma_{\mathrm{eff}}}
\def\sigv{\langle\sigma v\rangle}

\def\nuflub8{\phi^\nu_B}
\def\nuflube7{\phi^\nu_{Be}}

\def\rhodm{\rho_{\chi}}
\def\rdm{r_{\chi}}
\def\mdm{m_{\chi}}
\def\ldm{l_{\chi}}
\def\lesim{\lesssim}
\def\grsim{\gtrsim}
\def\vstar{v_{*}}
\def\Vstar{V_{*}}
\def\vmean{\bar{v}}
\def\Rstar{R_{*}}
\def\Lstar{L_{*}}
\def\rhostar{\rho_{*}}
\def\Mstar{M_{*}}

\def\ndm{n_{\chi}}

\def\Ldm{L_{\chi}}
\def\Lmod{L_{mod}}
\def\Lobs{L_{obs}}
\def\Pobs{P_{obs}}
\def\Pmod{P_{mod}}
\def\Msun{M_{\bigodot}}
\def\Rsun{R_{\bigodot}}
\def\Lsun{L_{\bigodot}}

\def\aap{Astron.\ Astrophys.\ }
\def\apj{Astrophys.\ J.\ }
\def\apjl{Astrophys.\ J.\ Lett.\ }

\def\aj{Astron.\ J.\ }
\def\mnras{Mon.\ Not.\ Roy.\ Astron.\ Soc.\ }
\def\physrep{Phys.\ Rept.\ }
\def\prd{Phys.\ Rev.\ D\ }

\def\araa{Annu.\ Rev.\ Astron.\ Astrophys.\ }
\def\jcap{J.\ Cosmol.\ Astropart.\ Phys.\ }

\def\pasp{Publications\ of\ the\ Astronomical\ Society\ of\ the\ Pacific}

\def\km{\,\mathrm{km}}

\def\gev{\,\mathrm{GeV}}
\def\mev{\,\mathrm{MeV}}

\def\cm{\,\mathrm{cm}}
\def\s{\,\mathrm{s}}

\newcolumntype{p}{D{,}{\pm}{-1}}

\def\c{\,\mathrm{c}}
\def\erg{\,\mathrm{erg}}
\def\g{\,\mathrm{g}}

\begin{document}



\title{Constrain the Dark Matter Electron Cross Section from Pulsating White Dwarfs}

\author{Jia-Shu Niu}
\email{jsniu@itp.ac.cn}
\affiliation{Key Laboratory of Theoretical Physics, Institute of Theoretical Physics, Chinese Academy of Sciences, Beijing, 100190, P.R.China}
\affiliation{School of Physical Sciences, University of Chinese Academy of Sciences, No.19A Yuquan Road, Beijing 100049, P.R.China}
\author{Weikai Zong}
\affiliation{Astronomy Department, Beijing Normal University,    Beijing 100875, P.R.China}
\author{Hui-Fang Xue}
\affiliation{Astronomy Department, Beijing Normal University,    Beijing 100875, P.R.China}

\date{\today}

\begin{abstract}
We propose a novel and feasible method to detect dark matter (DM) electron interaction via pulsating white dwarfs (WDs) in the central region of globular clusters. Annihilation of the DM particles captured by those WDs can provide additional energy source along the natural cooling evolution of WDs and the cooling print can be well offered by precise asteroseismology. The measurement of the long time scale physical quantity -- the rates of period variation of pulsation modes -- could be used to constrain the cross section between DM particles and electrons ($\sige$), when DM particle mass $\mdm \grsim 5 \gev$. We construct estimations to prove that this original method is feasible and can be implemented in the challenging time-series photometry in the near future.
\end{abstract}

\pacs{}
\maketitle

\textit{Introduction.}---Although dark matter (DM) contributes to $26.8 \%$ of the total energy density of the Universe \citep{Plank2014}, the particle nature of DM remains largely unknown. The most popular candidates are weakly interacting massive particles (WIMPs) for which are currently searched with different strategies, i.e., direct detection, indirect detection and detection on large terrestrial colliders (see, e.g., \citep{Bertone2005} for reviews). The most stringent constraints from experiments of cross section are currently, for proton, $\sigp \lesim 1.1 \times 10^{-46} \cm^{2}$ at a WIMP mass of $\mdm = 50 \gev/\c^{-2}$, and for electron, $\sige < 3 \times 10^{-38} \cm^{2}$ at $\mdm = 100 \mev $ and $\sige < 10^{-37} \cm^{2}$ when $20 \mev \le \mdm \le 1 \gev$ \citep{LUX2017,XENON2012}.

An alternative search for DM probe can be upon on celestial objects, such as stars, that have huge volume and large mass, comparing to the manual facilities. There are some branches collaborators who had began working on: 
the Sun and main-sequence stars, considered the precise properties of their interior structures from helioseismology and asteroseismology (see, e.g., \citep{Frandsen2010,Iocco2012,Vincent2015}); and the compact stars (i.e., white dwarfs and neutron stars), considered their deficiency of nuclear energy source (see, e.g., \citep{Moskalenko2007,Bertone2008,Hurst2015,Amaro2016}). Nevertheless, all these attempts are attracted on the interactions between WIMPs and nucleons whose cross section ($\sigp$) has been constrained below $1.1 \times 10^{-46} \cm^{2}$ by direct detection \citep{LUX2017}. In order to enhance the interacting effects between WIMPs and nucleus over the entire stars, a circumstance with high DM local density is preferred, where the stars in such as galaxy center, globular cluster and dwarf galaxy can be a reasonable solution.

In this letter, we consider the hydrogen-atmosphere pulsating white dwarf stars (WDs) in the central region of globular clusters such as $\omega$ Cen, and focus on the DM-electron interactions. The hydrogen-atmosphere (DA) WDs consist of $\sim 80 \%$ WDs sample that are the ending fates of $\sim 98 \%$ stars in the Galaxy. Precise asteroseismology on DA variables (DAVs) can reveal their interior structures and determine the secular variations of period variation of oscillation modes spanning over long time scales (see, e.g., \citep{Winget2008,Fontaine2008} for reviews). The secular rates can be used to determine the evolutionary cooling rate of WDs. However, the DM captured by WDs in a local dense DM environment possibly affects the cooling rates of WDs predicted by standard cooling model (SCM). In reverse, we can use the observed WD cooling rates to constrain the cross section in a condition with known DM density. Moreover, WDs are the most electron-dense objects and can be  the best laboratory to measure the DM-electron cross section ($\sige$).

\textit{The capture rate of WIMPs in WDs.}---Galactic WIMPs are inevitably streaming through any celestial object. Those particles will loose energy when they scatter with nucleons (which we did not mainly considered in this letter) or electrons inside the celestial object, leading their speed down. If the velocity of the WIMP reaches below the escape velocity, the WIMP will be "captured", i.e. it becomes bound to the star. Regardless of the effect of evaporation which is not important in this letter where we consider the WIMPs mass $m_{\chi} > 5 \gev$ \citep{Gould1987b,Gould1990a}, the evolution of the total number of WIMPs, $N_{\chi}$, inside the star (or any celestial object) can be written as, 

\begin{equation}
\dot{N_{\chi}}=\Gamma_{c}- 2 \Gamma_{a},
\label{eq_capture}
\end{equation}
where $\Gamma_{c}$ is the particle capture rate, $\Gamma_{a}=\frac{1}{2} C_{a} N_{\chi}^2$ is the annihilation rate. Therefore, we have $N_{\chi} = \Gamma_{c} \tau \tanh(\frac{t}{\tau})$
with the equilibrium timescaletime $\tau = \sqrt{\frac{1}{C_{a} \Gamma_{c}}}$. When the dynamic equilibrium state reached, the WIMPs capture rate is balanced by the annihilation one \citep{Griest1987}, i.e., $\Gamma_{c} = 2 \Gamma_{a}$.

Considering the limits that almost all WIMPs crossing the star to be  captured, the WIMP capture rate is determined by the geometrical limit $\pi R_{\star}^{2}$ rather than the total interaction cross section $\sige N_{e}$. We thus use a effective cross section $\sigeff$ defined as

\begin{equation}
  \label{eq:sigeff}
\sigeff = \min \left[ \sige, \frac{\pi R_{\star}^2}{N_{e}} \right],
\end{equation}
where $N_{e}$ is the total number of electrons in the WD.

The capture rate for a  Maxwell-Boltzmann WIMP velocity distribution (in the observer's frame) by a star moving with an arbitrary velocity $\vstar$ relative to the observer is given by \citet{Gould1987a}, but we  did some additional simplifications as well: (i) a uniform distribution of matters in a WD: $\rhostar (r) = \rhostar = \Mstar/\Vstar$ ($\rhostar$, $\Mstar$ and $\Vstar$ are the density, mass and volume of a WD); (ii) the same chemical composition over the entire scattering volume $\Vstar$; (iii) a uniform temperature profile (calculated from Eq. \ref{eq:evelope_structure}) in WD because of the extremely high thermal conductivity of a electron degenerate core; (iv) as WDs are always electrically neutral， we use the value $\frac{1}{2} \frac{\Mstar}{m_{p}}$ and $\frac{1}{2} \frac{\rhostar}{m_{p}}$ as the total number of electrons ($N_{e}$)  and local number density of electrons ($n_{e}$) in a WD, respectively ($m_{p}$ and $m_{e}$ are the mass of proton and electron). With these preparations, we can formulate the capture rate in a WD as  
\begin{equation}
  \Gamma_{c}=  \int_{0}^{R_*} \diff r\ 4 \pi r^2 \frac{\diff \Gamma_{c}(r)}{\diff V},
  \label{eq:capture}
\end{equation}
where
\begin{eqnarray}
\frac{\diff \Gamma_{c}(r)}{\diff V} &=&  \left(\frac{6}{\pi}\right)^{1/2}
\sigeff \frac{\rhostar}{2 m_{p}}\frac{\rhodm}{\mdm}
\frac{v^{2}(r)}{\vmean^{2}} \frac{\vmean}{2 \eta A^2} \\
\nonumber &\times & \left\{ \left( A_+ A_- -\frac{1}{2}\right)
[\chi(-\eta,\eta)-\chi(A_-,A_+) ] \right.\\ \nonumber &+& \left.
\frac{1}{2} A_+ e^{-A_-^2} -\frac{1}{2} A_- e^{-A_+^2} -\frac{1}{2}
\eta e^{-\eta^2} \right\}
\label{eq:dcdv}
\end{eqnarray}
$$ A^2=\frac{3 v^2(r)\mu}{2 \vmean^2 \mu_-^2} \mbox{,  }\hspace{0.5cm}
A_{\pm}=A \pm \eta \mbox{,}
\hspace{0.5cm}\eta^2=\frac{\vstar^2}{2\vmean^2}$$
$$\chi(a,b)=\frac{\sqrt{\pi}}{2}[\mbox{Erf}(b)-\mbox{Erf}(a)]=\int_a^bdy e^{-y^2}$$
$$\mu_-=(\mu-1)/2 \mbox{,} \hspace{0.5cm} \mu=m_{\chi}/m_{e}$$
, where $\mdm$  and $\rhodm$ are respectively the WIMP mass and the WIMPs density at the star position. $\vmean$ is the WIMPs' speed dispersion of Maxwell-Boltzmann velocity distribution; the velocity of the star moving through the DM halo or sub-halo, labeled as $\vstar$.

The escape velocity at a given radius $r$ inside a star is given by
\begin{equation}
  \label{eq:esc_v}
  v(r) = \left[ 2G \int_{r}^{\infty} \diff r' \frac{M(r')}{r'^{2}}   \right]^{1/2} = \left[ \frac{G \Mstar}{\Rstar} \left( 3 - \frac{r^{2}}{\Rstar^{2}} \right) \right]^{1/2},
\end{equation}
where $M(r')$ denotes the total mass included in $r'$.

The Knudsen number, $K$, indicates the ''localization'' of the WIMPs transport:
\begin{equation}
  \label{eq:Knumber}
  K = \frac{\ldm (0)}{\rdm},
\end{equation}
with $\ldm (0)=\left[ \sige \cdot n_{e}(0) \right]^{-1} = \left[ \sige \cdot n_{e}\right]^{-1}$ is the mean free length in the center of the star and $\rdm = \sqrt{\frac{3 k T_{c}}{2 \pi G \rho_{c} \mdm}}$ is the typical scale of the DM core in the star ($T_{c}$ and $\rho_{c}$ are the temperature and density of the star's core).

Following \citet{Griest1987,Scott2009,Taoso2010}, in the case of large $K$ (for WDs), the WIMPs' distribution in star can be described by
\begin{equation}
\ndm (r)=\ndm (0) \cdot \exp \left[ - (\frac{r}{\rdm})^2  \right].
\label{eq:dis_K}
\end{equation}

The annihilation term can be computed by a separated way as follows:
\begin{equation}
\Gamma_{a} = \int_{0}^{\Rstar}  \diff r \ 4 \pi r^{2} \cdot \frac{1}{2} \sigv \ndm^{2} (r).
\label{eq:ann}
\end{equation}
The factor 1/2 (1/4) in the equation above is appropriate for self (not self) conjugate particles
and $\sigv$ the velocity-averaged annihilation cross-section.

If an equilibrium between capture and annihilation is reached the annihilation rate reduces to $\Gamma_{a} = 1/2 \Gamma_{c} $ and it is independent on the annihilation cross-section. With the value of $\sigv \simeq 3 \times 10^{-26} \ \cm^{3} \s^{-1}$, we can impose  $\Gamma_{a} = 1/2 \Gamma_{c} $  to do the normalization and get $\ndm(0)$. Thus, the distribution $\ndm (r)$ is specified and all the related values in this equilibrium state are known.

\textit{The rate of period variation of DAVs}---The period variation of a DAV is related to two physical processes in the star, the cooling of the star and the contraction of its atmosphere, 

\begin{equation}
  \frac{\dot P}{P} \simeq -a \frac{\dot T_{c}}{T_{c}} + b \frac{\dot R}{R},
  \label{eq:var_period}
\end{equation}
where $P$ is the pulsation period for the $m=0$ multiplet component, $T_{c}$ is the maximum (normally, core) temperature, $R$ is the stellar radius, and $\dot P$, $\dot T$ and $\dot R$ are the respective temporal variation rates \citep{Winget2008}. The constants $a$ and $b$ are positive numbers of order unity.

From the structure of a WD's envelope, we get:

\begin{equation}
  \label{eq:evelope_structure}
T_{0}^{\frac{7}{2}} = B \frac{L/L_{\odot}}{M/M_{\odot}},
\end{equation}
where $B \simeq 1.67 \times 10^{27}$ is a constant and $T_{0}$ is the interface temperature between the core and envelope.

If we  use the approximation $T_{0} \simeq T_{c}$ (for DAVs) in Eq. \ref{eq:evelope_structure},  substitute the result into Eq. \ref{eq:var_period} and ignore the mass variation term during the cooling, we can get:
\begin{equation}
  \label{eq:p_l}
\frac{\dot P}{P} \simeq - \frac{2a}{7} \frac{\dot L}{L}.
\end{equation}

According to the annihilation of WIMPs in a WD, if the equilibrium state has reached, the luminosity of the WD should be $\Lobs = \Lmod + \Ldm$, where $\Lobs$ is the total observed luminosity of the WD, $\Lmod$ is the normal luminosity in the SCM and $\Ldm$ is the luminosity purely due to the accretion and subsequent annihilation of WIMPs.

From previous section, we get $\Ldm = 2 \cdot \Gamma_{a} \mdm \c^{2} = \Gamma_{c} \mdm \c^{2}$.
One should note that once the equilibrium state reached, $\Ldm$ should not change with time. As a result, $\dot{L}_{obs} = \dot{L}_{mod}$. Thus, the relationships between the rate of period variation and the luminosity should be
\begin{equation}
  \label{eq:L_P_result}
  \frac{\dot{P}_{obs}}{\dot{P}_{mod}} = \frac{\Lmod}{\Lmod + \Ldm},
\end{equation}
in which we imposed that the period from model calculation $\Pmod$ equals the value from observation $\Pobs$.

The evolutionary rate of cooling of a WD depends on the stellar mass and core composition, and can be expressed as a function of the mean atomic weight $A$ \citep{Mestel1952, Kawaler1986, Kepler1995}
\begin{equation}
  \label{eq:dpdt}
  \dot{P}_{mod} = \diff P / \diff t = (3-4) \times 10^{-15} \frac{A}{14}\  \s \ \s^{-1}.
\end{equation}
In this letter, we use a mean atomic weight 14, which is consistent with the cooling rate of DAVs with a carbon-oxygen core \citep{Mukadam2013}.

\textit{Estimation.}---We perform our estimation in the case that a DAV in the central region of globular cluster. Considering the DAVs G117-B15A as an example to do estimation, the details of its structure are determined by asteroseismology from \citet{Romero2012,Corsico2012}, and we list in Tab. \ref{astero_results}.

\begin{table}[htb]
\begin{center}
\begin{tabular}{cc}
  \hline\hline
Quantity  &Asteroseismological model\\
  \hline
$\Mstar/\Msun$          &$0.593 \pm 0.007$                         \\
$\log(\Rstar/\Rsun)$          &$-1.882 \pm 0.029$                         \\
$\log(\Lstar/\Lsun)$          &$-2.497 \pm 0.030$                         \\
  $M_{He}/\Mstar$          &$2.39 \times 10^{-2}$                         \\
  $M_{H}/\Mstar$          &$(1.25 \pm 0.7) \times 10^{-6}$                         \\
  $X_{C}, X_{O}$ (center)          &$0.28^{+0.22}_{-0.09}, 0.70^{+0.09}_{-0.22}$                         \\

  \hline\hline
\end{tabular}
\end{center}
\caption{$\Msun$, $\Rsun$ and $\Lsun$ are the solar mass, radius and luminosity; $M_{H}$, $M_{He}$ are the mass of element hydrogen and helium in the star; $X_{C}$ and $X_{O}$ are the mass fractions of element carbon and oxygen in the star.}
\label{astero_results}
\end{table}

Using Eq. \ref{eq:sigeff} we got $\sigeff \simeq 7.5 \times 10^{-39}\ \cm^2$, which is more than an order of magnitudes lower than the current upper limit on $\sige$ (Here we assume the $10^{-37} \cm^2$ upper limit is also available for $\mdm \grsim 5 \gev$).  This value implies that almost all WIMPs crossing the star are captured by WD as long as we choose $\sige \grsim \sigeff$.

Imposing $\rdm < \Rstar$, we get another lower limit of WIMPs' mass ($\mdm \grsim 0.025 \gev$) which we considered in this letter. Combined with the previous one, the mass region which we considered in this letter is $\mdm \grsim 5 \gev$.

The WIMPs local density of where the WD locates would influence the luminosity of WIMPs' annihilation obviously. In order to get large WIMPs density, we choose $\omega$ Cen (a globular cluster) to be the environment to do the estimation. \citet{Amaro2016} has estimated $\rhodm \simeq 4 \times 10^{3} \gev \cm^{-3} $ near the center of $\omega$ Cen without an intermediate-mass black hole (IMBH) and $4 \times 10^{3} \gev \cm^{-3} \lesim \rhodm \lesim 4 \times 10^{9} \gev \cm^{-3}$ with an IMBH. We here choose $\rhodm = 4 \times 10^{3} \gev \cm^{-3} $ to do the estimation.

From \citet{Merritt1997,Van2006}, the members of this cluster are orbiting the center of mass with a peak velocity dispersion  $\vstar \simeq 7.9 \km \s^{-1}$. Near a IMBH, where orbital motion around a single mass dominates, the test particle (WIMP or star) velocities are Keplerian, $\vstar = \vmean$.

With the configurations above, we get the DM luminosity of the WD as $\Ldm \simeq 3.2 \times 10^{30} \erg \s^{-1}$, which is of the same order of $\Lmod \simeq 1.2 \times 10^{30} \erg \s^{-1}$. As a result, the period variation of WD should be $\dot{P}_{obs} \simeq 0.27 \times \dot{P}_{mod} \simeq (0.81-1.08) \times 10^{-15} \frac{A}{14} \s \s^{-1} \simeq (0.81-1.08) \times 10^{-15} \s \s^{-1}$, which is obviously smaller than the value from SCM \citep{Winget2008}.

\begin{figure*}
\centering
\includegraphics[width=0.8\textwidth,height=0.4\textheight]{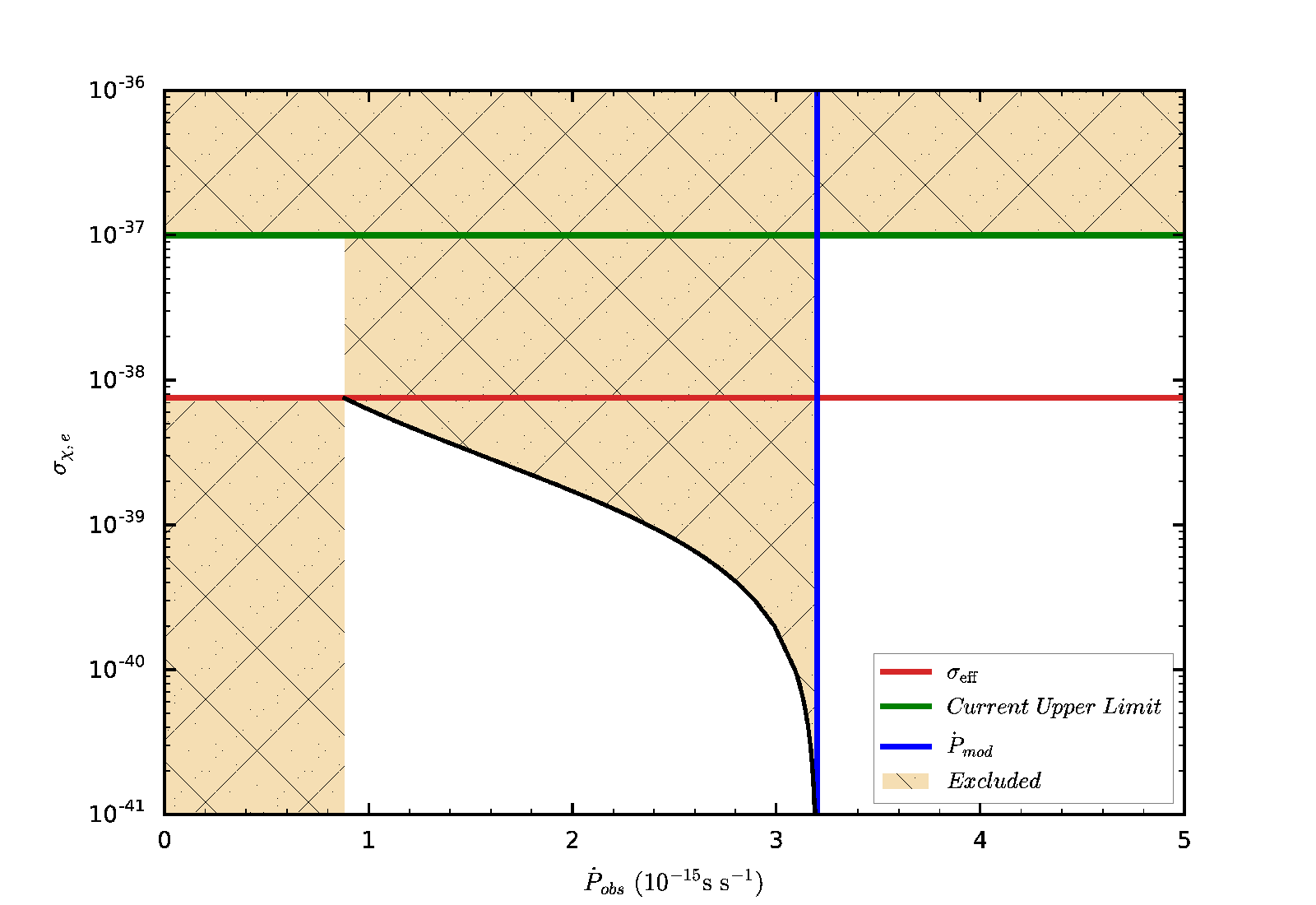}
\caption{The excluded region of $\sige$ with observed value of $\dot{P}_{obs}$. Colored region with slash lines indicate the excluded region with specific $\dot{P}_{obs}$. Colored lines indicate some important boundaries and threshold. The region $\dot{P}_{obs} > \dot{P}_{mod}$ is not considered in this letter, which we plot it for convenient. Note that $\dot{P}_{mod}$ is the asymptotic line of the black exclusive line. Considering the systematics of $\dot{P}_{mod}$, the exclusive line always stopped $< \dot{P}_{mod}$ in real cases.}
\label{fig:dp_cs}
\end{figure*}

With  $\dot{P}_{mod} = 3.2 \times 10^{-15} \s \s^{-1}$ \citep{Corsico2012}, we can present the constraint of $\sige$ by  $\dot{P}_{obs}$ in Fig. \ref{fig:dp_cs}: (a) if $\dot{P}_{obs} < 0.88 \times 10^{-15} \s \s^{-1}$ (depending on the geometrical limit of the star), it gives a lower limit to $\sige$; (b) if $0.88 \times 10^{-15} \s \s^{-1} \leq \dot{P}_{obs} \leq \dot{P}_{mod}$, it gives an upper limit to $\sige$; (c) if $\dot{P}_{obs} > \dot{P}_{mod}$, there should exist other unknown cooling mechanism beyond the SCM. If we include this new cooling mechanism into the SCM and update the value of $\dot{P}_{mod}$, the conclusions are all available in (a) and (b). We note that in region (b), the real  constraint of $\sige$ depends on the systematics of $\dot{P}_{mod}$. In this case, if we have more precise and reliable SEM and pulsation model for DAVs, we could get more stringent constraint on $\sige$ theoretically.

Moreover, we got $\tau \simeq 8.8\ yr$, which is really a short period compared with the time scale of  WD's formation process. Thus, we can consider WDs are always in the state of equilibrium. The total mass of WIMPs in the WD is about $2 \times 10^{18} \g \ll \Mstar $, then  its gravitational effects on the WD's interior structure can be ignored.

We also did a similar estimation for WIMPs which captured by the DAVs G117-B15A, according to the interactions with nucleus. In this case, we got $\Ldm \simeq 1.7 \times 10^{24} \erg \s^{-1}$ (here we choose $\sigp = 1.1 \times 10^{-46} \cm^{2}$), which can be ignored compared  with $\Ldm \simeq 3.2 \times 10^{30}$ from electrons. Consequently, it is a reasonable assumption to consider the WIMPs captured by interactions with electrons alone.

\textit{Discussion.}---In this letter we propose a novel and feasible scenario to measure the DM-electron cross section, a technique independent to the mass of DM particle if $\mdm \grsim 5 \gev$. The scenario is based on the assumption that DAV stars are found in a local dense DM environment region such as the central region of globular cluster. According to the measurement of the secular rates of the pulsation periods in those DAVs, we can constrain the DM-electron cross section.

\begin{figure*}
\centering
\includegraphics[width=0.8\textwidth,height=0.4\textheight]{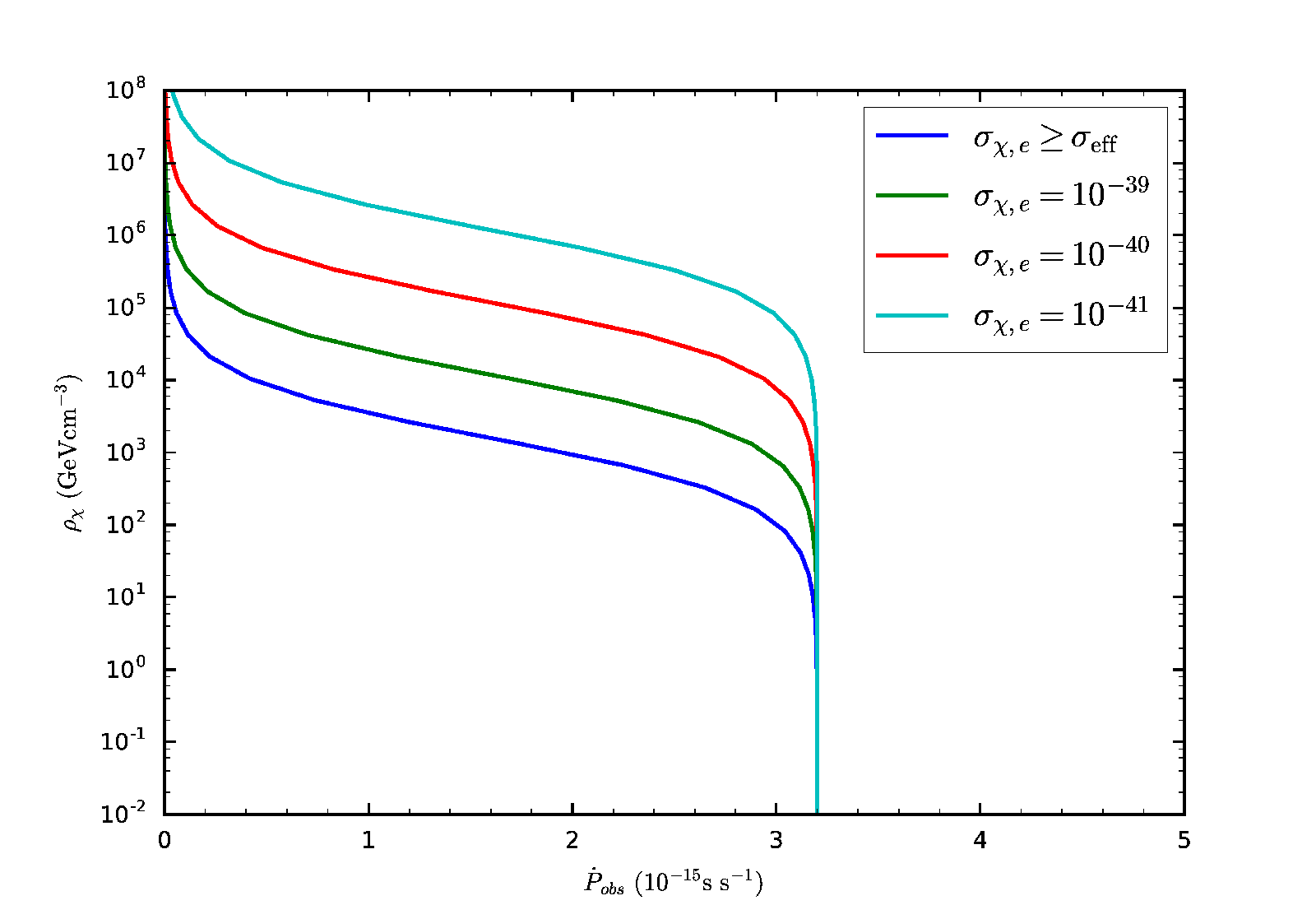}
\caption{The relation between $\dot{P}_{obs}$ and WIMPs local density $\rhodm$, with different $\sige$. Note that if $\sige \geq \sigeff$, the $\sige$ do not influence the profile which takes the same curve as $\sige = \sigeff$. }
\label{fig:dp_rhodm}
\end{figure*}

At the same time, if we give a specific value of $\sige$, in this scenario, the observed quantity $\dot{P}_{obs}$ could offer us the WIMPs local density in the position of the DAV. Fig. \ref{fig:dp_rhodm} represents the different profiles of $\rhodm$ vs. $\dot{P}_{obs}$ with different values of $\sige$. Additionally, ensemble of these DAVs can be used to establish the density profiles in these regions.

From our estimation, we note that once we detect the period variation of such DAVs, it can give us useful information, including: (a) if the rate of period variation is smaller than the threshold ($\dot{P}_{gl}$) which depend on the geometrical limits of the star (in this our estimation, $\dot{P}_{gl} \simeq 0.88 \times 10^{-15} \s \s^{-1}$), it could provide a lower limit on $\sige$; (b) if $\dot{P}_{gl} < \dot{P}_{obs} < \dot{P}_{mod}$, it could provide a upper limit on $\sige$; (c) if $\dot{P}_{obs} \grsim \dot{P}_{mod}$, it should exist some other unknown cooling mechanism, like axions (see, e.g., \citep{Corsico2012,Corsico2012b,Corsico2016,Battich2016}).

This scenario can be implemented by high-precision photometry from space mission such as HST, {\sl Kepler} and TESS in the near future.

We thank Tianjun Li and Jun Guo for useful discussion. This research was supported in part by the Natural Science Foundation of China under grant numbers 11135003, 11275246, and 11475238 (TL).

%

\end{document}